\documentclass[final]{agujournal2019}
\usepackage{url} 
\usepackage{lineno}

\usepackage{setspace}
\usepackage{threeparttable}
\usepackage{soul}
\usepackage{xcolor}
\usepackage{multirow}
\usepackage{CJK}

\journalname{GRL}
\graphicspath{{./}{figures/}}
\begin{document}
\title{The Fate of Simple Organics on Titan's Surface: A Theoretical Perspective}

\authors{Xinting Yu\affil{1}, Yue Yu\affil{2,3}, Julia Garver\affil{4}, Xi Zhang\affil{2}, Patricia McGuiggan\affil{5}}

\correspondingauthor{Xinting Yu}{xinting.yu@utsa.edu}

\affiliation{1}{Department of Physics and Astronomy, University of Texas at San Antonio, 1 UTSA Circle, San Antonio, TX 78249, USA}
\affiliation{2}{Department of Earth and Planetary Sciences, University of California Santa Cruz, 1156 High St, Santa Cruz, CA 95064, USA}
\affiliation{3}{Department of Astronomy, University of Geneva}
\affiliation{4}{Department of Physics, University of California Santa Cruz, 1156 High St, Santa Cruz, CA 95064, USA}
\affiliation{5}{Department of Materials Science and Engineering, Johns Hopkins University, 3400 N. Charles Street, Baltimore, MD 21218, USA}


\begin{keypoints}
\item Most simple organics land as solids on Titan's surface, including all nitriles, triple-bonded hydrocarbons, and benzene.
\item Organics may float on Titan's lakes via porosity or capillary force-induced flotation, the latter is unique to HCN ice on ethane-rich lakes.
\item Porosity-induced flotation of millimeter-sized and larger particles may explain the transient radar-bright magic islands on Titan's lakes.

\end{keypoints}

\begin{abstract}
Atmospheric photochemistry on Titan continuously transforms methane and nitrogen gases into various organic compounds. This study explores the fate of these molecules when they land on Titan's surface. Our analytical exploration reveals that most simple organics found in Titan's atmosphere, including all nitriles, triple-bonded hydrocarbons, and benzene, land as solids. Only a few compounds are in the liquid phase, while only ethylene remains gaseous. For the simple organics that land as solids, we further examine their interactions with Titan's lake liquids. Utilizing principles of buoyancy, we found that flotation can be achieved via porosity-induced (25-60\% porosity) or capillary force-induced buoyancy for HCN ices on ethane-rich lakes. Otherwise, these ices would sink and become lakebed sediments. By evaluating the timescale of flotation, our findings suggest that porosity-induced flotation of millimeter-sized and larger sediments is the only plausible mechanism for floating solids to explain the transient ``magic islands" phenomena on Titan's lakes.
\end{abstract}

\section*{Plain Language Summary}

Titan, Saturn's largest moon, has a unique atmosphere that transforms simple gases like methane and nitrogen into more complex organic compounds. In this study, we explored what happens to these organic compounds when they reach Titan's surface. We find that most molecules would land as solids. We also looked at what happens when these solids land on Titan's hydrocarbon lakes. Imagine a sponge, full of holes; if the solids are like this, with 25-60\% of their volume being empty space, they can float. Some solids, like hydrogen cyanide ice, can also float due to surface tension effects. If these conditions are not met, they sink into the lake liquids, adding to the lakebed sediments. We examine whether floating rafts can explain a mysterious feature on Titan's lakes known as the ``magic islands". These are temporary bright spots seen by radar. By looking at how long the materials will float for each scenario, our study suggests that the magic islands might be made of large chunks of porous organic solids.

\section{Introduction}

Titan's thick methane (CH$_4$) and nitrogen (N$_2$) atmosphere has enabled rich photochemistry to occur in its upper atmosphere. Photochemistry leads to the creation of a myriad of organic molecules, with at least 17 gas-phase simple organic species already identified in Titan's atmosphere \cite{2022..xxx..xxxA}. The increasing complexity of the photochemical reactions eventually leads to the formation of complex refractory organic particles, which constitute Titan's characteristic haze layers \cite{2005Natur.438..765T}. Following their formation, these organic molecules, being heavier than the background N$_2$-CH$_4$ atmosphere, would descend through Titan's atmosphere. The sharp temperature drop in Titan's stratosphere enables these molecules to condense into liquids or ices, forming stratospheric clouds \cite{1984Icar...59..133S, 2018SSRv..214..125A}. Ultimately, these organics deposit onto Titan's surface, either onto dry areas or into Titan's lakes and seas.

The importance of understanding the fate of simple organics on Titan's surface is underscored by NASA's forthcoming Dragonfly mission, due to arrive on Titan in 2034. The mission will primarily explore Titan's surface material in the equatorial region, where dry surfaces dominate. Thus, it is crucial to identify the range of materials likely to be found there and their associated phases. The species that are deposited on the surface of Titan, existing as solids or liquids, could further interact to form ``cryominerals" such as co-crystals, a topic of ongoing research \cite{cable2021titan}. Future co-crystal investigations could take combinations of the molecules that can remain as solids/liquids on Titan's surface as experimental candidates.

Throughout the course of the Cassini mission, the two most intriguing puzzles about Titan's lakes and seas are: 1) the striking smoothness of the lake liquids, with wave heights typically less than a few millimeters \cite{2009GeoRL..3616201W,2014GeoRL..41..308Z,2017E&PSL.474...20G, 2019NatAs...3..535M}, activities (possibly waves) were only identified in a few circumstances \cite{2014PlSci...3....3B, 2014NatGe...7..493H, 2016Icar..271..338H}, and 2) the transient ``magic islands"  features, observed as radar-bright features on Titan’s two largest seas, Ligeia Mare \cite{2014NatGe...7..493H, 2016Icar..271..338H} and Kraken Mare \cite{2018LPI....49.2065H}. The lack of surface roughness has been attributed to either the lack of winds in the polar regions of Titan \cite{2010Icar..207..932L} or a floating layer of sedimented materials on Titan’s lake liquids \cite{2019NatGe..12..315C}. Several plausible hypotheses have been proposed for the ``magic islands"  phenomena, involving wind-generated waves, suspended solids, floating solids, or nitrogen gas bubbling \cite{2014NatGe...7..493H, 2017NatAs...1E.102C, 2017Icar..289...94M, 2018ApJ...859...26C, 2019GeoRL..4613658F}. 

To explore these phenomena, interactions between deposited solid materials and Titan's lake liquids warrant further investigation. These solids, possibly simple organic ices or complex organic hazes, might reach Titan's lakes through precipitation, fluvial/aeolian delivery, and runoffs from nearby islands. The interactions between the haze particles and the lake liquids have been investigated in \citeA{2020ApJ...905...88Y} and \citeA{2022PSJ.....3....2L}. The simple organics in the solid phase could also potentially constitute the floating materials observed on Titan's lake liquids. Accordingly, an in-depth floatability study of various simple organics on Titan's lake liquids is necessary. In this work, we examine the solid-liquid interactions and floatability of various materials on Titan. Assessing the interplay between these organics and Titan's lake liquids could provide insights into the overall smoothness of the lakes and the nature of the transient ``magic islands" phenomena.

\section{The Fate of Simple Organics Landing ``Dry" Surfaces of Titan}
Using phase change points and saturation vapor pressures (SVPs) of detected gas-phase simple organics species in an updated Titan material property database \cite{2022..xxx..xxxA}, we determine the physical phase of 17 hydrocarbon and nitrile species detected in Titan's atmosphere. These species include methane (CH$_4$), ethane (C$_2$H$_6$), ethylene (C$_2$H$_4$), acetylene (C$_2$H$_2$), propane (C$_3$H$_8$), propene (C$_3$H$_6$), allene (C$_3$H$_4$-a), propyne (C$_3$H$_4$-p), diacetylene (C$_4$H$_2$), benzene (C$_6$H$_6$), hydrogen cyanide (HCN), cyanoacetylene (HC$_3$N), acetonitrile (CH$_3$CN), propionitrile (C$_2$H$_5$CN), acrylonitrile (C$_2$H$_3$CN), cyanogen (C$_2$N$_2$), and dicyanoacetylene (C$_4$N$_2$). 

\begin{figure}[h]
\centering
\includegraphics[width=0.8\textwidth]{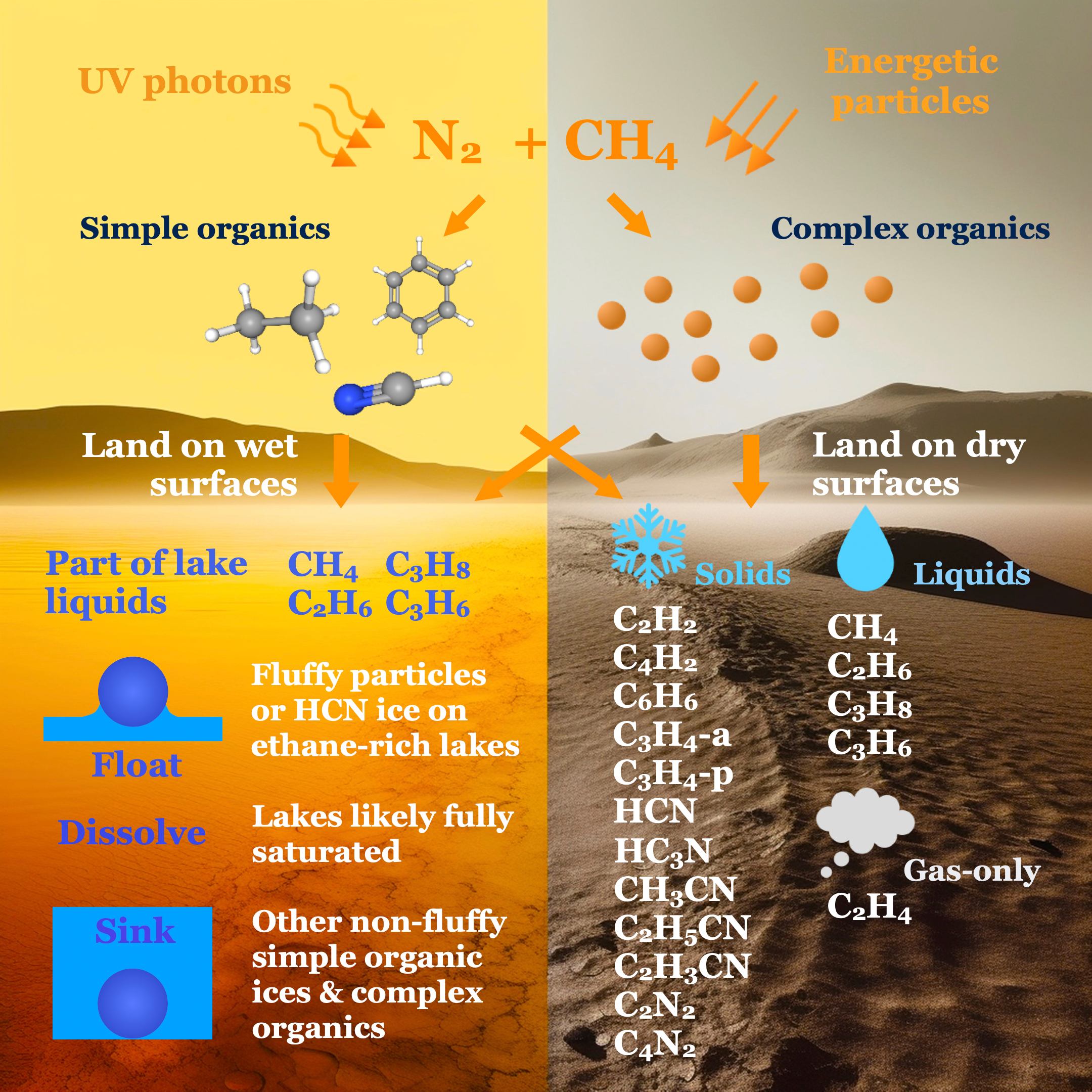}
\caption{Summary of the fate of simple and complex organics on Titan's surface (background image AI generated by X. Yu using Midjourney).}
\label{fig:summary}
\end{figure}

In Figure \ref{fig:summary}, we detail the phase states of simple organics upon reaching Titan's surface. Most of the higher-order hydrocarbon species and all the nitrile species remain in the solid phase upon deposition on Titan's surface due to the high triple point temperatures of these species (see SI Table S1). This group includes twelve out of the seventeen species included in this study: triple-bonded hydrocarbons (C$_2$H$_2$, C$_4$H$_2$, C$_3$H$_4$-a, C$_3$H$_4$-p), C$_6$H$_6$, and all nitriles (HCN, HC$_3$N, CH$_3$CN, C$_2$H$_5$CN, C$_2$H$_3$CN, C$_2$N$_2$, and C$_4$N$_2$). Alkanes and alkenes like C$_3$H$_8$ and C$_3$H$_6$ would transition to the liquid phase upon reaching Titan's surface. CH$_4$ and C$_2$H$_6$ predominantly transform into liquids unless the surface temperature dips below their respective triple points. The only species expected to stay exclusively gaseous on Titan's surface is C$_2$H$_4$, due to strong photochemical sinks in the lower stratosphere limiting its condensation \cite{2016ApJ...829...79W}. Our simple estimation aligns with the findings of photochemical models incorporating condensation schemes, \citeA<for HCN in>{2011Icar..215..732L}, \citeA<for the nitrile species in>{2014Icar..236...83K, 2016ApJ...829...79W}, and \citeA<all the major hydrocarbon and nitrile species in>{2019Icar..324..120V}. Future work that incorporates more updated SVPs from \citeA{2022..xxx..xxxA} into a sophisticated photochemical model could facilitate the estimation of the deposition rate of each condensate on Titan's surface.

\section{The Fate of Simple Organics Landing ``Wet" Surfaces of Titan}

The lake liquids on Titan are primarily composed of a ternary mixture of non-polar liquids: methane, ethane, and nitrogen \cite{2016AREPS..44...57H}. C$_2$H$_4$ is the sole species anticipated to be only in the gas phase upon reaching Titan's surface. C$_3$H$_8$ and C$_3$H$_6$ would manifest as rainfall on Titan's surface. Given their miscibility with lake liquids \cite{1987AdSpR...7e..71R, 2013GeCoA.115..217G}, they will be mixed into the N$_2$-CH$_4$-C$_2$H$_6$ ternary mixture, becoming minor components of the lake liquids. The fate of the twelve solid-phase species upon reaching Titan's surface is dependent on their solubility in the lake liquids. Previous solubility studies \cite{1987AdSpR...7e..71R, 2009ApJ...707L.128C, 2013ApJ...768L..23C, 2014Icar..242...74M} suggest that except for C$_2$H$_2$, the other eleven solid-phase species have limited solubility in Titan's ternary lake mixture, likely leading to saturation over time. The saturation level of C$_2$H$_2$ is subject to debate \cite{1987AdSpR...7e..71R, 2009ApJ...707L.128C}, but for our analysis, we assume it too reaches saturation upon contact with the lakes. We recognize, however, that dynamic lake movements, including currents and precipitation, may affect the dissolution and distribution of compounds like C$_2$H$_2$. However, in this theoretical framework, we assume all compounds would remain solid when they are in contact with the lake liquids, either floating or becoming lakebed sediments.

\subsection{The Porosity Consideration}

\begin{figure}
    \includegraphics[width=\textwidth]{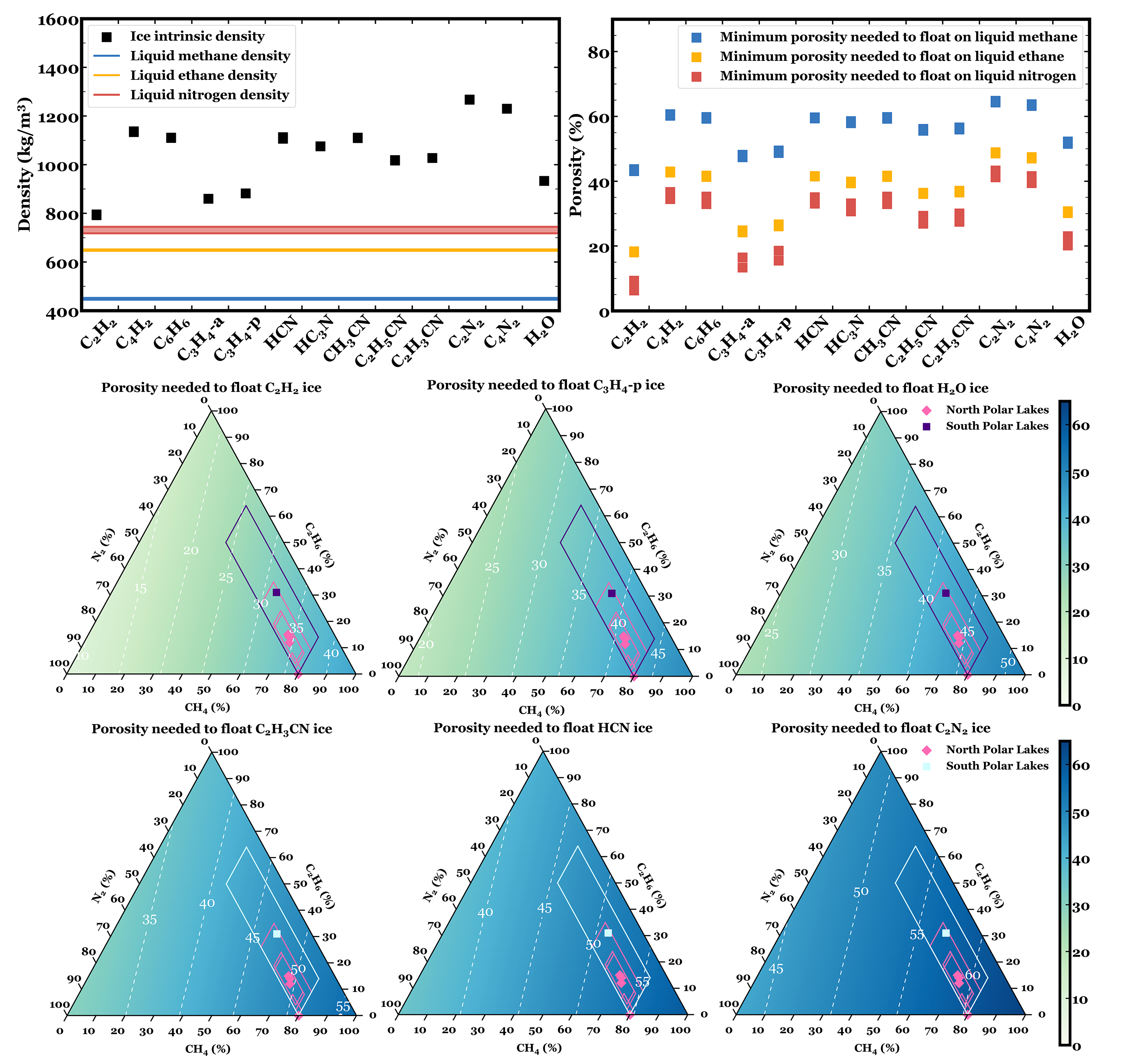}
    \caption{Top-left: Intrinsic densities of the 12 simple organic ices and water ice at Titan's surface temperature (blue squares) and liquid density values for methane, ethane, and nitrogen  (blue, orange, and red lines). Top-right: Minimum porosity required for these species to float on pure liquid methane, ethane, or nitrogen (colored squares). The density variation and the threshold porosity values between 90-95 K are smaller than the symbol sizes. Middle and bottom panel: ternary diagrams depicting the porosity levels required for selected ices to float on the ternary mixture. The light blue and pink diamond symbols indicate observed compositions of Titan's northern and southern polar lakes, with diamond-shaped boxes showing the uncertainty \cite{2020JGRE..12506558P}.} 
    \label{fig:density}
\end{figure}

To determine if simple organic ices would float or sink into Titan's lakes, we first compare their density to the lake liquids. Examining the intrinsic density (the maximum density without porosity) of the ices shows that all twelve species and water ice have higher intrinsic densities than each component of the N$_2$-CH$_4$-C$_2$H$_6$ ternary mixture, see Figure \ref{fig:density}(top-left).

However, if the ice contains pore space, its average density will be lower than its intrinsic density. When the average density equals the liquid density, additional porosity will cause flotation of the particle. Figure \ref{fig:density}(top-right) outlines the minimum porosity needed for flotation for each lake component. We also evaluate the minimum porosity required for these ices to float on the liquid mixture. Results for selected ices are shown in the middle and bottom panels of Figure \ref{fig:density}. Because Titan's lakes are mostly methane-rich, as indicated by the existing Cassini RADAR loss-tangent measurements \cite{2016JGRE..121..233L, 2016ITGRS..54.5646M, 2018Icar..300..203M, 2018E&PSL.496...89M, 2019NatAs...3..535M, 2020JGRE..12506558P} and phase equilibrium theoretical predictions \cite{2013Icar..222...53T, 2013GeCoA.115..217G}, the possible liquid compositions only cover a fraction of the ternary diagram. Since the density of liquid methane is the lowest among the ternary mixture, the required porosity level for flotation on Titan's lakes is relatively high, with lower-order alkynes (C$_2$H$_2$, C$_3$H$_4$-a, C$_3$H$_4$-p) needing the least porosity, around 25\%-35\%. Higher-order hydrocarbons, water ice, and all nitrile species need higher porosity, up to 50\% for C$_2$N$_2$ ice.

To date, there are no laboratory measurements investigating the potential porosity levels these ices can achieve on Titan. Existing laboratory data on crystalline ices indicate porosity values from a few percent to $\sim$35 \% \cite{2008P&SS...56.1748S, 2017Icar..296..179S, 2022..xxx..xxxA}. Amorphous ices can exhibit higher levels of porosity, up to 40-50 \% \cite{2022..xxx..xxxA}. However, the surface temperature of Titan is too warm for amorphous ice formation. Nonetheless, these laboratory ice analogs are typically produced from gas-phase vapor deposition on a cooled substrate within a vacuum chamber, which may result in different morphologies and pore structures compared to snowfalls/sediments on Titan. Fresh terrestrial snow is another potential analog to the snowfall/sediment particles on Titan and it typically has high porosity (40-50 \%, up to 90 \%) \cite<e.g.,>{2008BoLMe.126..249C, zermatten2014tomography, fu2019analysis}. More laboratory experiments are required to investigate porosity for these ices under Titan-relevant conditions. If Titan's ice particles have porosity closer to terrestrial snow, they could possess a high enough porosity to trigger flotation. Given that the porosity threshold is lowest for the low-order alkynes, they are more likely to float on Titan's lakes.

\subsection{The Contact Angle Consideration}

\begin{figure}
    \centering
    \includegraphics[width=\textwidth]{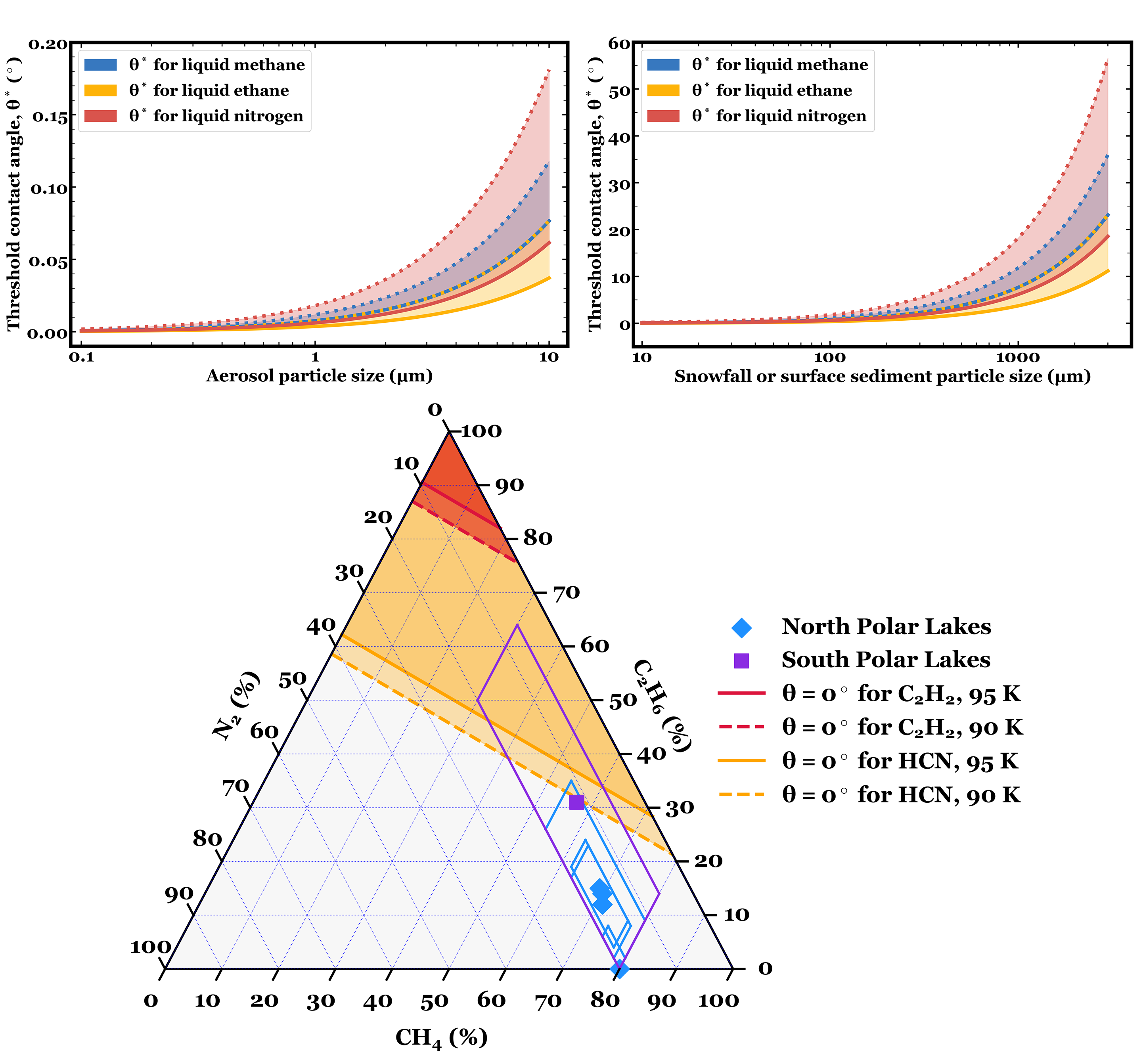}
    \caption{Top: threshold contact angles to trigger ice flotation through capillary force, plotted against particle size, considering: 1) aerosol-sized particles, 0.1 - 10 $\mu$m, and 2) snowfall and sand-sized surface sediments, 10 $\mu$m-3 mm. Shaded areas indicate threshold contact angle ranges for methane, ethane, and nitrogen (blue, orange, and red shaded areas). Bottom: CH$_4$-C$_2$H$_6$-N$_2$ ternary diagram showing zones with non-zero contact angles for C$_2$H$_2$ (red) and HCN (orange). Boundaries for zero contact angles at 90 K and 95 K are marked by dashed and solid lines. Square symbols and diamond boxes denote the observed composition and uncertainty of measured Titan lakes.}
    \label{fig:theta}
\end{figure}

Given that the twelve simple organic ice species on Titan's surface possess larger intrinsic density than the lake liquids, and assuming their porosity isn't sufficient to warrant flotation, it is necessary to examine the impact of capillary force. Capillary-force triggered flotation depends primarily on the contact angle between the ice particles and the lake liquids. Figure \ref{fig:theta}(top) shows the minimum contact angles needed for flotation across plausible particle sizes on Titan. For aerosol-sized particles (0.1-10 $\mu$m, \citeA{2008P&SS...56..669T}), the density of the particle has minimal effect on the threshold contact angle values. Only a slight above zero contact angle is needed to trigger flotation due to capillary forces. Meanwhile, particles with sizes like snowfall or aeolian surface sediments (10 $\mu$m to $\sim$3 mm, \citeA{2006Icar..182..230B, 2006Sci...312..724L, 2018JGRE..123.2310Y}) require higher contact angles to offset gravity, as this threshold contact angle is proportional to particle size. Therefore, it is easier for smaller particles to float on Titan's lakes due to capillary forces.

Using the wetting theory, we computed the contact angles between the twelve simple organic ices and the three primary constituents of the lake liquids. For liquid nitrogen/methane, because of their low surface tensions, they have zero contact angles with all simple organic ices. Thus, any ice particles would be completely wetted by these two liquids, leading to zero capillary forces. Conversely, pure liquid ethane, being the liquid component with the highest surface tension, has zero contact angles with most simple organic species, with the exceptions of C$_2$H$_2$ and HCN, which have non-zero contact angles of approximately 24.0$^\circ$ and 49.7$^\circ$, respectively. 

Thus, the more ethane is incorporated into the lake liquids, the higher the surface tension of the liquid mixture, making the liquid mixture less wetting. In Figure \ref{fig:theta}(bottom), we highlighted areas where non-zero contact angles are formed on a ternary diagram, which are for C$_2$H$_2$ (in red) and HCN (in orange) ices. Given that most lakes on Titan are methane-rich, there is no overlap between the non-zero contact angle zone for C$_2$H$_2$ and the known lake composition zone, suggesting C$_2$H$_2$ is unlikely to float on Titan's lakes through the surface tension effect. For HCN ice, the non-zero contact angle zone does have some overlaps with the uncertainty of the measured lake compositions, when ethane's fraction is high ($>30\%$). Specifically, the non-zero contact angle only overlaps with the uncertainty of Ontario Lacus's composition, the only lake with measured composition in the south polar regions of Titan. Lakes in the north polar regions with measured compositions all have zero contact angle with HCN. These results suggest it is unlikely that capillary forces would cause simple organic species to float in any methane-rich lakes on Titan. However, for larger particles, a higher than zero threshold contact angle is required (see Figure \ref{fig:theta}(top)), making it more difficult for these ices to float via capillary forces on Titan's lake liquids.

The bulk liquids in Titan's large lakes and seas should be typically well-mixed given that the surface temperature on Titan should be well above 86 K \cite{2020PSJ.....1...26S}. However, the topmost surface layer of the liquid mixture may exhibit lower surface tension than the bulk mixture, due to preferential concentration of the component with lower surface tension at or near the surface of the mixture \cite{poling2001properties}, making the liquid more wetting at the surface level. Thus, the non-zero contact angle will further reduce compared to in Figure \ref{fig:theta}(bottom), making capillary force-induced floating more challenging to realize on Titan's lakes.

\section{Implications on the Titan Lake Radar Observations}

Among all the hypotheses that have been proposed to explain the smoothness of the lakes and the observed radar-bright transient features on Titan, two hypotheses can explain both observations simultaneously: the absence/presence of waves or floating solids. The lack of wind-generated waves might account for the general smoothness of the lakes, while occasional waves could give rise to transient radar-bright ``magic island" phenomena. Likewise, a uniformly thin layer of floating solids could also explain the overall smoothness, with large clusters of floating solids potentially visible as ``magic islands". 

One way to assess whether floating solids cause the transient radar-bright features is to consider the residence time of the solids. Existing observations suggest that these features persist for a couple of hours to several weeks \cite{2014GeoRL..41.1432M, 2016JGRE..121..233L}. If the observations in \citeA{2018LPI....49.2065H} are indeed examining the same transient feature, the floating solids would need to remain afloat for at least 2 hours. Understanding the fate of the floating solids --- assuming they initially float due to high porosity or high contact angle --- could provide insight into whether particle flotation can cause the ``magic islands" on Titan. 

\begin{figure}
\centering\includegraphics[width=0.7\textwidth]{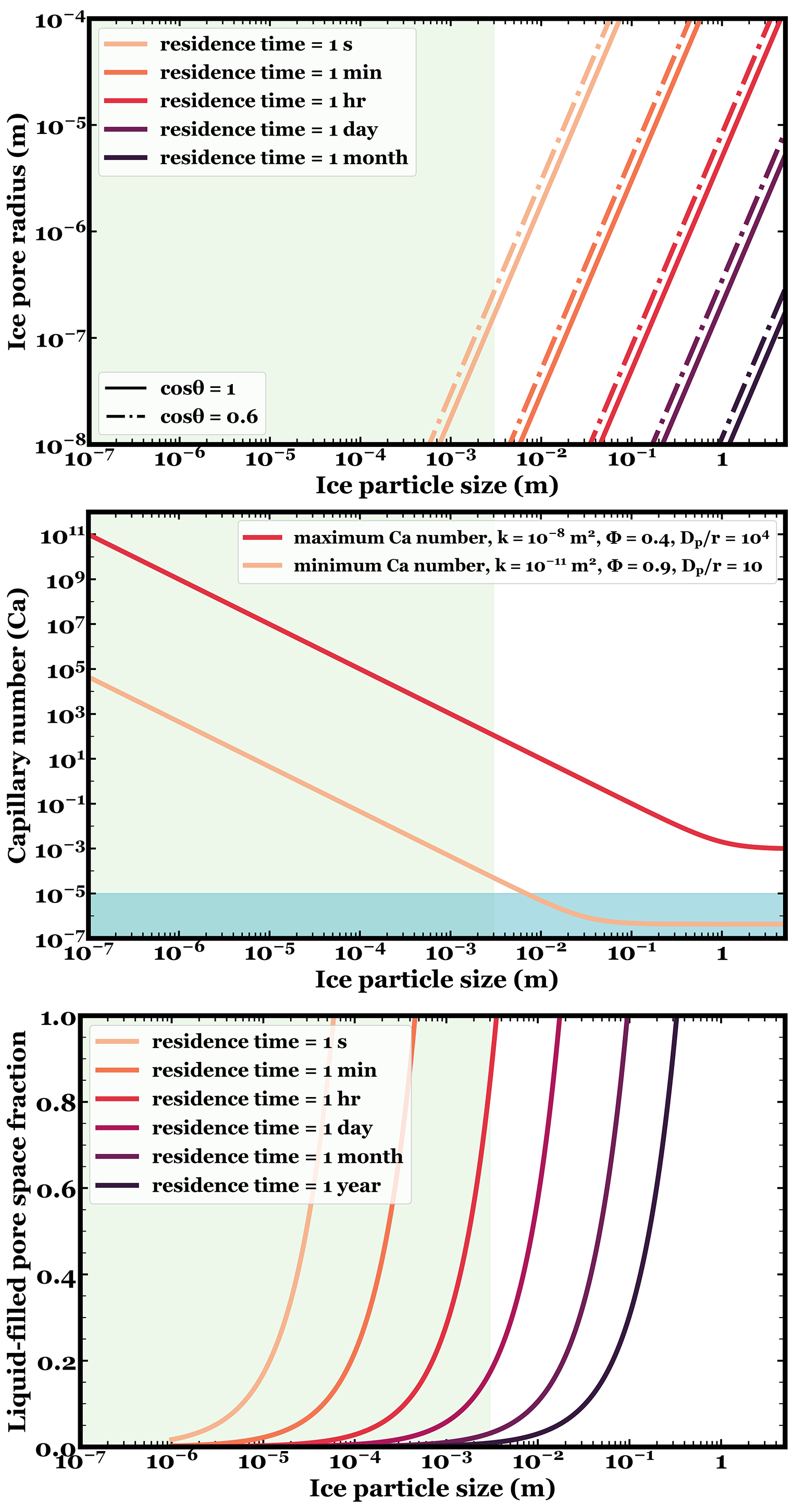}
\caption{Top: Wetting timescale of ice grains against particle size and pore radius calculated by the Washburn Equation, with the solid and dash lines for $\cos\theta =1$ and $\cos\theta =0.6$. Middle: Ice Capillary number, Ca, by particle size, depicting the maximum and minimum Ca number (red and orange lines) considering uncertain ice particle properties. The blue shaded area indicates Ca$<10^{-5}$, where gas trapping starts to occur. Bottom: Ice grain residence time against particle size and liquid-filled pore fraction calculated by the diffusion equation. The green shaded areas show possible particle sizes of aerosols, snowfall particles, and sand-sized sediments.}
\label{fig:time}
\end{figure}

In the event that the porosity of the simple organic ices is responsible for their flotation, we can expect that, over time, lake liquids would gradually seep in and fill any pore spaces that are connected to the exterior, leading to the eventual sinking of the ice particle. We can use the Lucas-Washburn equation \cite{1921PhRv...17..273W, lucas1918ueber} to estimate the wetting timescale ($t$) of a porous particle with particle size D$_p$:
\begin{equation}
    t = \frac{2D_p^2\eta}{\gamma_lr\cos\theta},
    \label{eq:washburn}
\end{equation}
where $\eta$ is the dynamic viscosity of the liquid (for liquid methane, $\eta=1.8\times$10$^{-4}$ Pa$\cdot$s at 94 K, \citeA{1967Phy....33..547B, 1973Phy....70..410H}), $r$ is the pore radius of ice particle (r/D$_p$ $\ll$ 1). This simple estimation yields a wetting time less than one second in most scenarios for sand-sized and smaller particles, see Figure \ref{fig:time}(top). Even though some simple organic ice species (HCN and C$_2$H$_2$) have higher contact angles with the lake liquids, the timescale will not increase by orders of magnitude. Therefore, for floating substances to persist for longer periods, there needs to be a constant supply of floating materials for the entire period. Or the ice particles would have to possess sufficient closed pores that cannot be penetrated by the lake liquids. Yet, the latter is unlikely for two reasons: 1) terrestrial snow particles, a possible analog to Titan's organic ices, have a very low closed pore to open pore ratio ($<0.01$) \cite{2012TCry....6..939C}, suggesting that the pore space is primarily open. While the structure of Titan's snowfall/surface sediment particles is unknown, it is unlikely for them to possess high porosity and a high fraction of closed pores at the same time. 2) There is evidence that Titan's lakebed is largely comprised of insoluble solid hydrocarbon/nitrile materials \cite{2016JGRE..121..233L}, implying that most of the snowfall material would eventually reach the bottom of the lakes and seas, while the latter scenario will lead to long-term flotation without sinking.

A phenomenon on Earth that defies the Washburn equation is the long flotation time of pumice rafts. Pumices can float on the oceans on Earth for days and weeks, some even for months to years before sinking \cite{1958DSR.....5...29R, kent1984drift}, while possessing all the similar traits to simple organic ices versus liquid hydrocarbons: 1) pumice have a much higher intrinsic density than water ($\sim$2500 kg/m$^3$ versus 1000 kg/m$^3$), 2) being very wettable by water, and 3) have mostly open, connected pores \cite{1986BVol...48..209W}. A possible explanation for the long flotation time of pumice is that a significant volume fraction of gases is trapped in pore space during liquid infiltration \cite{2017E&PSL.460...50F}. In this case, the physical process that dictates the flotation timescale of pumice is gas diffusion instead of water infiltration. Gas trapping likely occurs when the Capillary number ($Ca$), the ratio between the viscous force and the capillary force becomes very small \cite{lenormand1984role}, or Ca $<10^{-5}$. The middle panel of Figure \ref{fig:time} shows the Capillary number calculated given the range of parameters that could exist on Titan. When gas trapping occurs, a very wetting liquid would still cover all surface area of the particle quickly, leaving the center of pores or chains of pores open. Here we found the minimum particle size to trigger gas trapping on Titan is 5 mm and above. This means the flotation timescale for sand-sized and smaller particles is governed by liquid penetration, while gas diffusion dominates the flotation timescale for coarser particles. The gas diffusion timescale can be approximated as \cite{2017E&PSL.460...50F}:
\begin{equation}
    t = \frac{D_p^2}{D\xi^2},
\label{eq:diffusion}
\end{equation}
where $D$ is the diffusion coefficient of the gas through the liquid, and $\xi$ is the fraction of pore space filled by the liquid. Figure \ref{fig:time} bottom panel shows the flotation timescale if gas diffusion is the dominant process that governs particle buoyancy through time. Given the same particle size, the flotation timescale is orders of magnitude longer for gas diffusion than for wetting/liquid penetration.

In summary, if the simple organic ice particles float on Titan's lakes due to high porosity, the flotation timescale would be short ($<1$ min) for particles smaller than a few millimeters but could be much longer ($>$ 1 Earth day) for larger particles. If floating materials are indeed behind Titan's ``magic island" phenomena, they must be coarse particles, several millimeters in size or larger. Although a full exploration of all geological processes capable of producing these coarse materials is beyond the scope of this study, runoff from nearby terrains or activities like transient undersea cryovolcanism (e.g., \citeA{2018SciA....4.1121C}) could be plausible sources. Conversely, finer particles like aerosols, snowfalls, and sand-sized sediments are unlikely to maintain prolonged flotation. Thus, while the continual deposition of atmospheric aerosol particles could contribute to the smoothness of Titan's lake surfaces, they are unlikely to form the more substantial ``magic islands".

For the simple organic ices floating on Titan's lakes due to large capillary forces, the flotation would last for an extended period, persisting until the particles are entirely dissolved. However, capillary force-induced flotation seems unlikely to cause the lake observations on Titan for two primary reasons: 1) it is improbable in methane-rich northern polar lakes, where the transient features are observed; 2) if it does occur, flotation would persist and particles may never sink to the bottom of the lakes. 

Note that the above theoretical framework primarily applies to a stagnant lake surface. While dynamic factors like currents and waves could influence the flotation timescales of particles, but given the characteristically calm and smooth nature of Titan's lakes and seas \cite{2009GeoRL..3616201W,2014GeoRL..41..308Z,2017E&PSL.474...20G, 2019NatAs...3..535M}, their impact is expected to be less significant compared to the more dynamic aquatic environments on Earth.

\section{Conclusion}

In our theoretical examination of the fate of simple organics on Titan's surface, guided by the data from \citeA{2022..xxx..xxxA}, we draw the following conclusions:
\begin{itemize}
\item Most simple organics will land as solids on Titan's surface, while methane, ethane, propane, and propene will be liquid. Ethylene will be gaseous only.
\item Simple organics may achieve buoyancy on Titan's lake liquids through porosity-induced or capillary force-induced flotation, with porosity-induced flotation being feasible if the simple organics on Titan resemble terrestrial snow. Capillary force-induced flotation is only viable for HCN ice in ethane-rich lakes.
\item Porosity-induced flotation of millimeter-sized and larger particles is the only plausible mechanism for floating solids to explain Titan's transient radar-bright magic islands. Shorter or longer timescales for other flotation mechanisms do not align with the observations.
\end{itemize}

\section{Acknowledgements}
X. Yu and J. Garver are supported by the 51 Pegasi b Postdoctoral Fellowship from the Heising-Simons Foundation. X. Yu is also supported by the NASA Cassini Data Analysis Program Grants 80NSSC21K0528 and 80NSSC22K0633 and the NASA Planetary Science Early Career Award 80NSSC23K1108. X. Zhang is supported by the NASA Solar System Workings Grant 80NSSC19K0791, NASA Interdisciplinary Consortia for Astrobiology Research (ICAR) grant 80NSSC21K0597, and the NSF Astronomy and Astrophysics Research Grants 2307463. Y. Yu thanks the Other Worlds Laboratory at UCSC for summer research support. We thank Alan Whittington for his valuable insights regarding transient cryovolcanism.

\section{Open Research}

The data used to perform this study is available at \citeA{2022..xxx..xxxAA}.

\nocite{2012P&SS...60...62C, 2005Natur.438..785F, 2008GeoRL..35.9204H, himmelblau1964diffusion, 2016Icar..277..103H, 2016ApJ...816L..17J, 2019ApJ...877L...8J, kaelble1970dispersion, owens1969estimation, rabel1971einige, richardson2018experimental, 2019Icar..317..337T, tyn1975diffusion, wilke1955correlation}

\bibliography{reference.bib}

\end{document}